\documentclass[doublecol]{epl2}

\usepackage{amsfonts}
\usepackage{amsmath}
\usepackage{amssymb}
\usepackage{booktabs}
\usepackage{hyperref}
\usepackage{tabularx}

\hypersetup{
  colorlinks=true,linkcolor=blue,citecolor=blue,
  filecolor=blue,urlcolor=blue,breaklinks=true
}

\newcommand{\otoprule}{\midrule[\heavyrulewidth]}

\title{Remarks on the Dirac oscillator in $(2+1)$ dimensions}

\author{Fabiano M. Andrade\inst{1} \and Edilberto O. Silva\inst{2}}
\shortauthor{Andrade and Silva}

\institute{
  \inst{1}
  Departamento de Matem\'{a}tica e Estat\'{i}stica,
  Universidade Estadual de Ponta Grossa -
  84030-900 Ponta Grossa-PR, Brazil
  \\
  \inst{2}
  Departamento de F\'{i}sica,
  Universidade Federal do  Maranh\~{a}o -
  Campus Universit\'{a}rio do Bacanga,
  65085-580 S\~{a}o Lu\'{i}s-MA, Brazil
}


\abstract{
In this work the Dirac oscillator in $(2+1)$ dimensions is
considered.
We solve the problem in polar coordinates and discuss the dependence of
the energy spectrum on the spin parameter $s$ and angular momentum
quantum number $m$.
Contrary to earlier attempts, we show that the
degeneracy of the energy spectrum can occur for all possible values of
$sm$.
In an additional analysis, we also show that an isolated bound state
solution, excluded from Sturm-Liouville problem, exists.
}

\date{\today}

\begin{document}

\maketitle

The Dirac oscillator, first introduced in \cite{NCA.1967.51.1119} and
after develop in \cite{JPA.1989.22.817}, has been an usual model for
studying the physical properties of physical systems in various branches
of physics.
In the context of theoretical contributions, the Dirac oscillator has
been analyzed under different aspects such as the study of the
covariance properties and Foldy-Wouthuysen and Cini-Touschek
transformations \cite{JPA.1989.22.821}, as a special case of a class of
chiral solutions to the automorphism gauge field equations
\cite{JMP.1993.34.4428}, and hidden supersymmetry produced by the
interaction $iM\omega\beta \mathbf{r}$,  where $M$ is the mass, $\omega$
the frequency of the oscillator  and  $\mathbf{r}$  is the position
vector,  when it plays a role of anomalous magnetic interaction
\cite{PRL.1990.64.1643} (see also
Refs. \cite{JPA.1989.22.821,EJP.1995.16.135}).

Recently, the one-dimensional Dirac oscillator has had its first
experimental realization \cite{PRL.2013.111.170405}, which made the
system more attractive from the point of view of applications. 
The Dirac oscillator in $(2+1)$ dimensions, when the third
spatial coordinate is absent,  has also been studied in
Refs. \cite{PRA.2008.77.033832,PRA.2007.76.041801,PRA.1994.49.586}. 
Additionally, this system was proposed in \cite{arXiv:1311.2021} to
describe some electronic properties of monolayer an bylayer graphene.
For a detailed approach of the Dirac oscillator see the
Refs. \cite{Book.1998.Strange,Book.Moshinsky.1996}.

In this Letter, we address the Dirac oscillator in $(2+1)$ dimensions.
In \cite{PRA.1994.49.586}, it was argued that the energy eigenvalues are
degenerated only for negative values of $k_{\vartheta}s$, where
$k_{\vartheta}$ represents the angular momentum quantum number and $s$
the spin projection parameter.
This result, however, is not correct, as properly shown in this work.
Additionally, an isolated bound state solution for the Dirac oscillator
in $(2+1)$ is worked out.

We begin by writing the Dirac equation in $(2+1)$ dimensions 
($\hbar=c=1$)
\begin{equation}\label{eq:dirac}
  \left(
    \beta \gamma \cdot \mathbf{p}+\beta M
  \right)\psi=E\psi,
\end{equation}
where $\mathbf{p}=(p_{x},p_{y})$ is the momentum operator and $\psi$ is
a two-component spinor.
The Dirac oscillator is obtained through the following nonminimal
substitution \cite{JPA.1989.22.817}:
\begin{equation}\label{eq:oscillator}
  \mathbf{p}\to\mathbf{p}-i M\omega \beta \mathbf{r},
\end{equation}
where $\mathbf{r}=(x,y)$ is the position vector and $\omega$ stands for
the Dirac oscillator frequency.
Thus, the relevant equation is
\begin{equation}\label{eq:dirac_oscillator}
 \left[
    \beta \gamma \cdot
    (\mathbf{p}-i M\omega \beta\mathbf{r})+\beta M
  \right]\psi=E\psi.
\end{equation}
In three dimensions the $\gamma$ matrices are conveniently defined in
terms of the Pauli matrices \cite{PRL.1990.64.503}
\begin{equation}
  \beta \gamma_{x}=\sigma_{x},  \qquad
  \beta \gamma_{y}=s\sigma_{y}, \qquad
  \beta =\sigma_{z},
\end{equation}
where $s$ is twice the spin value, with $s=+1$ for spin ``up'' and
$s=-1$ for spin ``down''.
In this manner, eq. \eqref{eq:dirac_oscillator} can be written as
\begin{equation}\label{eq:Hdirac_osc}
  \left(
    \sigma_{x}\pi_{x}+s\sigma_{y}\pi_{y}+\sigma_{z}M
  \right)
  \psi=E\psi, 
\end{equation}
where $\pi_{j}=p_{j}- i M\omega\sigma_{z} r_{j}$.

As usual, we write eq. \eqref{eq:Hdirac_osc} in polar coordinates
$(r,\phi)$
\begin{equation}\label{eq:diracopolar}
e^{is\sigma_{z}\phi}
  \left[
    \sigma_{x}\partial_{r}+
    \sigma_{y}\left(\frac{s}{r}\partial_{\phi}-iM\omega r\right)
  \right]\psi= i(E-M\sigma_{z})\psi.
\end{equation}
If one defines the spinor as
\begin{equation}
  \psi=
  \left(
    \begin{array}{c}
      \psi_{1} \\
      \psi_{2}
    \end{array}
  \right),
\end{equation}
eq. \eqref{eq:diracopolar} leads to
\begin{subequations}\label{eq:dirac_pi}
\begin{align}
  \label{eq:dirac_pi_chi1}
  \left(
    \partial_{r} - i\frac{s}{r} \partial_{\phi} -M\omega r
  \right)\psi_{2} = {} & i e^{is\phi} (E-M)\psi_{1},\\
  \label{eq:dirac_pi_chi2}
  \left(
    \partial_{r} + i\frac{s}{r} \partial_{\phi} + M\omega r
  \right)\psi_{1} = {} & ie^{-is\phi} (E+M)\psi_{2}.
\end{align}
\end{subequations}
We decomposes the spinor as
\begin{equation}\label{eq:ansatz}
  \psi=
  \left(
    \begin{array}{c}
      \psi_{1} \\
      \psi_{2}
    \end{array}
  \right)=
  \left(
    \begin{array}{c}
      \sum_{m}   f_{m}(r)\;e^{i m   \phi} \\
      \sum_{m} i g_{m}(r)\;e^{i(m+s)\phi}
    \end{array}
  \right),
\end{equation}
where $m = 0,\pm 1,\pm 2,\pm 3,\ldots$ is the angular momentum quantum 
number.
The factor $i$ on the lower spinor component is included for later
convenience.
By replacing eq. \eqref{eq:ansatz} into eq. \eqref{eq:dirac_pi}, we
can write the two coupled first-order radial equations
\begin{subequations}
  \label{eq:rad}
  \begin{align}
    \left[\frac{d}{dr}+\frac{s(m+s)}{r}- M\omega r\right]g_{m}(r)= {} &
    (E-M)f_{m}(r),\label{eq:radf}\\
    \left[-\frac{d}{dr}+\frac{sm}{r}- M\omega r\right]f_{m}(r)= {} & 
    (E+M)g_{m}(r).\label{eq:radg}
  \end{align}
\end{subequations}
Now the role of the $i$ in the lower component of eq. \eqref{eq:ansatz}
is apparent. 
It was inserted to ensure that the radial part of the spinors is
manifestly real.

The problem of the Dirac oscillator in $(2+1)$ dimensions represented by
the  eqs. \eqref{eq:radf} and \eqref{eq:radg} for $E\neq\pm M$ can be
mapped into a Sturm-Liouville problem for the upper and lower components
of the Dirac spinor.
In this manner, as we will show, the solutions can be found by solving a
Schr\"{o}dinger-like equation.
An isolated solution for the problem, excluded from the Sturm-Liouville
problem, can be obtained considering the particle at rest, i.e., 
$E = \pm M$ directly in the first order equations in \eqref{eq:radf} and
\eqref{eq:radg}.  
Such solution for the Dirac equation in $(1+1)$ dimensions was
investigated in Ref.
\cite{AoP.2013.338.278} (see also Refs. 
\cite{PS.2008.77.045007,IJMPE.2007.16.2998,IJMPE.2007.16.3002,
JPA.2007.40.263,PS.2007.75.170,PLA.2006.351.379}).
We are seeking for bound state solutions subjected to
the normalization condition
 \begin{equation}\label{eq:norm}
\int_{0}^{\infty} \left(|f_{m}(r)|^{2}+|g_{m}(r)|^{2}\right) r dr =1.
\end{equation}

Let us begin by determining the isolated bound states solutions.
So, for $E=-M$, we can write
\begin{subequations}\label{eq:radE=-M}
  \begin{align}
    \left[\frac{d}{dr}+\frac{s(m+s)}{r}- M\omega r \right]g_{m}(r)= {} &
    -2M f_{m}(r),\label{eq:radfE=-M}\\
    \left[-\frac{d}{dr}+\frac{sm}{r}- M\omega r\right]f_{m}(r)= {}  & 
    0,\label{eq:radgE=-M}
  \end{align}
\end{subequations}
whose general solutions are
\begin{subequations}\label{eq:fgE=M}
  \begin{align}
    f_{m}(r) = {} & a_{-} r^{sm} e^{-M\omega r^2/2},\label{eq:fE=-M}\\
    g_{m}(r) = {} & \left[b_{-} - 2 M a_{-} I_{-}(r) \right] 
    r^{-sm-1} e^{M\omega r^2/2},\label{eq:gE=-M}
  \end{align}
\end{subequations}
where $a_{-}$ and $b_{-}$ are constants.
In \eqref{eq:gE=-M}, $I_{-}(r)$ can be expressed in terms of the
upper incomplete Gamma function \cite{Book.1972.Abramowitz}
\begin{equation}
  \label{eq:incgamma}
  \Gamma(a,x)=\int_{x}^{\infty}t^{a-1}e^{-t}dt, \qquad \Re(a)>0.
\end{equation}
In fact,
\begin{equation}
  I_{-}(r) =
  \frac{\Gamma\left(s m+1,M\omega r^{2}\right)}{2 (M\omega)^{sm+1}}.
\end{equation}
As $M\omega>0$, there are no integer values for $sm$ that the functions
in  \eqref{eq:fE=-M} and \eqref{eq:gE=-M} are square-integrable.
Therefore, there is no bound state solution for $E=-M$.
In the other hand, for $E=M$, from \eqref{eq:rad} we can write
\begin{subequations}\label{eq:radE=M}
\begin{align}
\left[\frac{d}{dr}+\frac{s(m+s)}{r}- M\omega r \right]g_{m}(r)= {} &
0,\label{eq:radfE=M}\\
\left[-\frac{d}{dr}+\frac{sm}{r}- M\omega r\right]f_{m}(r)= {}  & 
2 M g_{m}(r),\label{eq:radgE=M}
\end{align}
\end{subequations}
whose general solutions are
\begin{subequations}\label{eq:fgE=M}
\begin{align}
  f_{m}(r) = {} & 
  \left[b_{+} - 2 M a_{+} I_{+}(r)\right] 
  r^{sm} e^{-M\omega r^2/2},\label{eq:fE=M}\\
  g_{m}(r) = {} & a_{+} r^{-1-sm} e^{M\omega r^2/2},\label{eq:gE=M}
\end{align}
\end{subequations}
where $a_{+}$ and $b_{+}$ are constants, and
\begin{equation}
  I_{+}(r) =
  \frac{(-M\omega)^{sm}}{2}\Gamma\left(-s m,-M\omega
    r^{2}\right).
\end{equation}
A normalizable solution requires $a_{+}=0$.
In this case, the function $f_{m}(r)$ is square-integrable only
for $sm\geq 0$. 
Therefore,
\begin{equation}
   \left(
    \begin{array}{c}
      f_{m}(r) \\
      g_{m}(r)
    \end{array}
  \right)
    = b_{+} r^{sm} e^{-M\omega r^2/2}
    \left(
    \begin{array}{c}
      1\\
      0
    \end{array}
    \right),
    \qquad sm \geq 0.
\end{equation}

Now, for $E\neq\pm M$, by manipulation of eqs. \eqref{eq:radf} and
\eqref{eq:radg}, we can decouple them and obtain the following
Schr\"odinger-like second order differential equations for the
components: 
\begin{multline}\label{eq:second}
\bigg\{\frac{d^{2}}{dr^{2}}+\frac{1}{r}\frac{d}{dr}-
\frac{[m+(s\mp s)/2]^2}{r^{2}}+2M\omega s[m+(s\mp s)/2]
\;\;\;\;\;\\
-M^{2}\omega^{2}r^{2}+(E^{2}-M^{2})\bigg\}
\left(\begin{array}{c}f_{m}(r)\\g_{m}(r)\end{array}\right)=0.
\end{multline}
Our task now is to solve eq. \eqref{eq:second}.
Using the change of variable $\rho=M\omega r^2$, eq. \eqref{eq:second}
for the $f_{m}(r)$ assumes the form
\begin{equation}\label{eq:ode_Rrho}
  \rho f_{m}''(\rho)+f_{m}'(\rho)
  -\left(
    \frac{m^2}{4\rho}+\frac{\rho}{4} -\frac{k^2}{4\gamma}
  \right)f_{m}(\rho)=0,
\end{equation}
with $\gamma=M\omega$ and 
\begin{equation}
k^{2}=E^{2}-M^{2}+2\gamma (sm+1).
\end{equation}
Studying the asymptotic limits of eq. \eqref{eq:ode_Rrho}, and the
finiteness at the origin leads us to the following solution:
\begin{equation}\label{eq:Mrho}
  f_{m}(\rho)=\rho^{|m|/2}e^{-\rho/2} w(\rho).
\end{equation}
Substitution of eq. \eqref{eq:Mrho} into eq. \eqref{eq:ode_Rrho},
results
\begin{equation}\label{eq:ode_Mrho}
  \rho w''(\rho)
  +\left(1+|m|-\rho\right) w'(\rho)
  -\left(
    \frac{1+|m|}{2}-\frac{k^2}{4\gamma}
  \right)w(\rho)=0.
\end{equation}
Equation \eqref{eq:ode_Mrho} is a confluent hypergeometric-like equation
\begin{equation}
z w''(z)+(b-z) w'(z)-a w(z)=0,
\end{equation}
where $w(z)$ is the confluent hypergeometric function
\cite{Book.1972.Abramowitz}, with $a={(1+|m|)}/{2}-{k^2}/{4\gamma}$ and
$b=1+|m|$.
In this manner, the only acceptable solution for eq. \eqref{eq:ode_Mrho} 
is the confluent hypergeometric function of the first kind
$_{1}F_{1}(a,b,z)$. 
As $b=1+|m| \geq 1$, the other linearly independent solution, the
confluent hypergeometric function of the second kind $U(a,b,z)$, is
rejected because it is irregular at the origin. 
Consequently,
\begin{align}\label{eq:general_sol_2d_dirac}
  f_{m}(\rho)
  =a_{m} \;\rho^{|m|/2} e^{-\rho/2}\;
  _{1}F_{1}\left(d,1+|m|,\rho\right),
\end{align}
where
\begin{equation}
d=\frac{1+|m|}{2}-\frac{k^2}{4 \gamma}.
\end{equation}

In order to find the energy spectrum of the Dirac oscillator, we should
establish as convergence criterion the condition $d=-n$, with $n$ a
nonnegative integer.
Therefore, the energy levels are given by
\begin{equation}\label{eq:energy_do}
E=\pm\sqrt{M^2+2M\omega(2n+|m|-sm)}, \quad n=0,1,2,\ldots,
\end{equation}
and in this case $_{1}F_{1}\left(-n,1+|m|,\rho\right)$ can be written in
terms of the generalized Laguerre polynomials
$L_{n}^{|m|}(\rho)$ \cite{Book.1972.Abramowitz}.
In this manner, the upper component of the bound state  wave function is 
\begin{align}\label{eq:eigenfunction_2d_dirac}
  f_{m}(r)
  =a_{m}\;
  \rho^{|m|/2} e^{-\rho/2}\;
  L_{n}^{|m|}(\rho).
\end{align}
The lower component is obtained in an analogous
manner by directly solving the eq. \eqref{eq:second} and the result is
\begin{align}\label{eq:eigenfunction_2d_dirac_g}
  g_{m}(r)
  =b_{m}\;
  \rho^{|m+s|/2} e^{-\rho/2}\;
  L_{n-s\Theta(ms)}^{|m+s|}(\rho),
\end{align}
where $\Theta(x)$ is the Heaviside function, and $a_{m}$ and $b_m$ are
constants subject to the normalization condition in \eqref{eq:norm}.

\begin{table}[t]
  \begin{center}
    \resizebox{\columnwidth}{!}{
    \begin{tabularx}{1.1\columnwidth}{cccccccccccc}
      \toprule
      $m$ $\Rightarrow$ & {\bf -5}&{\bf -4}&{\bf -3}&{\bf -2}&{\bf -1}&{\bf \,0}
      &{\bf \,1}&{\bf \,2}&{\bf \,3}& {\bf \,4}&{\bf \,5}\\\cmidrule(l){1-1}
      $n$ $\Downarrow$ & & & & & & & & & & &\\\otoprule
      {\bf 0}& 10 & 8 & 6 & 4 & 2 & 0 & 0 & 0 & 0 & 0 & 0 \\\midrule
      {\bf 1}&12 & 10 & 8 & 6 & 4 & 2 & 2 & 2 & 2 & 2 & 2 \\\midrule
      {\bf 2}&14 & 12 & 10 & 8 & 6 & 4 & 4 & 4 & 4 & 4 & 4 \\\midrule
      {\bf 3}&16 & 14 & 12 & 10 & 8 & 6 & 6 & 6 & 6 & 6 & 6 \\\midrule
      {\bf 4}&18 & 16 & 14 & 12 & 10 & 8 & 8 & 8 & 8 & 8 & 8 \\\midrule
      {\bf 5}&20 & 18 & 16 & 14 & 12 & 10 & 10 & 10 & 10 & 10 & 10 \\
      \bottomrule
    \end{tabularx}
  }
  \end{center}
  \caption{The energy spectrum for the Dirac oscillator in $(2+1)$
    dimensions as a function of the good quantum numbers $n$ and $m$ for
    $s=1$.  
    For convenience the values correspond to $(E^2-M^2)/2M\omega$.}
  \label{tab:tab1}
\end{table}

\begin{figure}
  \centering
  \includegraphics*[width=\columnwidth]{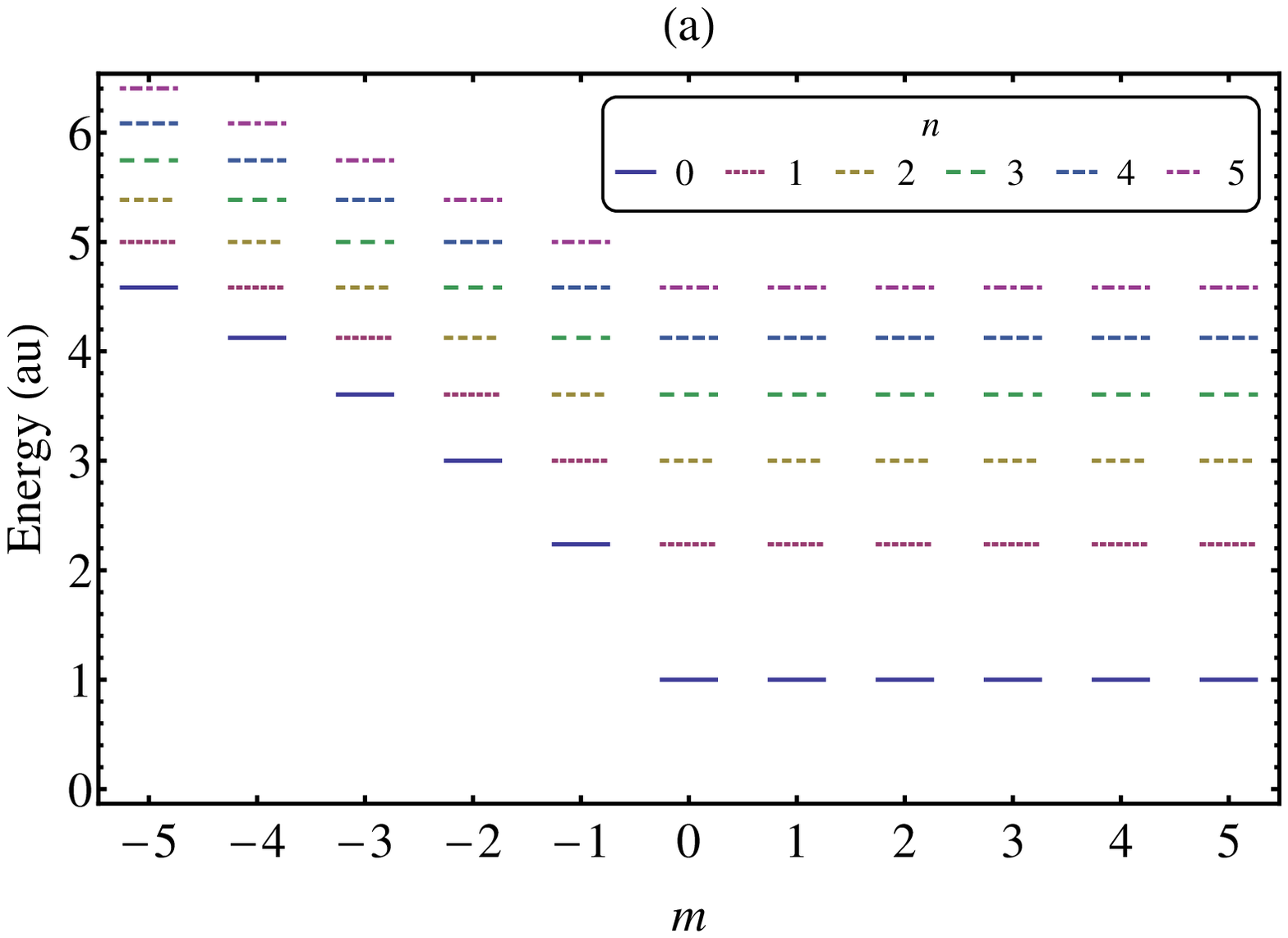}\vspace{0.5cm}
  \includegraphics*[width=\columnwidth]{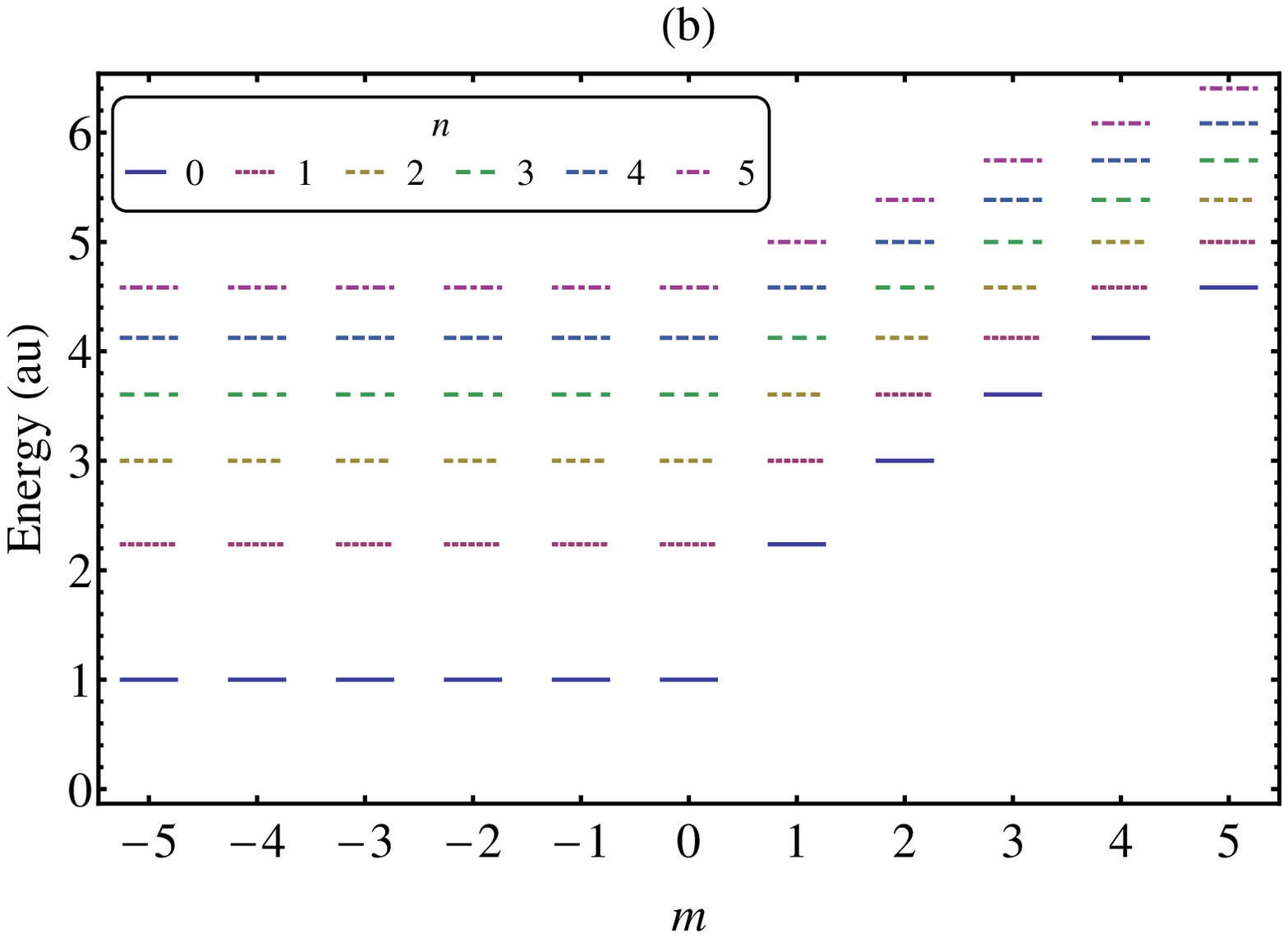}
  \caption{ \label{fig:fig1}
    (Color online) The positive energy spectrum,
    eq. \eqref{eq:energy_do}, for the Dirac oscillator
    in $(2+1)$ dimensions for different values of $n$ and $m$ with
    $M=\omega=1$ and for: (a) $s=1$ and (b) $s=-1$.
    Notice that levels with quantum numbers $n \pm q$  for $s=1$
    ($s=-1$) have the same energy as levels with $m \pm q$ ($m\mp q$),
    with $q$ an integer.
    For $sm \geq 0$, it is clear that the energy spectrum is independent of
    $m$.}
\end{figure}

It is important to note that, the energy levels expressed in
\eqref{eq:energy_do} are spin dependent and, contrary to the results of
\cite{PRA.1994.49.586}, the energy expression yield an infinity
degeneracy for all possible values of $sm$. 
Indeed, for $sm<0$, we have:
for $s=1$ and $m<0$ all levels with $n \pm q$ and $m \pm q$ have the
same energy, while for $s=-1$ and $m>0$ the equal energy levels are
those with $n \pm q$ and $m \mp q$, being $q$ an integer. 
Moreover, for $sm\geq 0$, all the energy levels are independent of the
angular quantum number $m$.
This behavior is depicted in Fig. \ref{fig:fig1} and represented in
Table \ref{tab:tab1}.

We can also obtain the energy spectrum for the Dirac oscillator in
$(2+1)$ dimensions following the standard derivation of Moshinsky and
Szczepaniak \cite{JPA.1989.22.817}.
Indeed, operating
$[\sigma_{x}\pi_{x}+s\sigma_{y}\pi_{y}+\sigma_{z}M +E]$ on 
eq. \eqref{eq:Hdirac_osc}, one obtains
\begin{equation}\label{eq:diracoquad}
  [p^2+M^{2}\omega^2r^2 -2M\omega(\sigma_{z} + sL_{z})]\psi=
  (E^2-M^{2})\psi,
\end{equation}
where $L_{z}=-i\partial_{\phi}$.
Equation \eqref{eq:diracoquad}, restoring the factors $\hbar$ and
$c$, in terms of components, becomes
\begin{subequations}\label{eq:2ddiracoscillatorcomp}
\begin{multline}
  2Mc^{2}\left(
    \frac{p^2}{2M}+\frac{1}{2}M\omega^2r^2-\hbar\omega-s\omega L_{z}
  \right)\psi_1=\\
  (E^2-M^{2}c^{4})\psi_1,
\end{multline}
\begin{multline}
  2Mc^{2}\left(
    \frac{p^2}{2M}+\frac{1}{2}M\omega^2r^2+\hbar\omega-s\omega L_{z}
  \right)\psi_2=\\
  (E^2-M^{2}c^{4})\psi_2.
\end{multline}
\end{subequations}
Equation \eqref{eq:2ddiracoscillatorcomp}, for $s=1$, agreed with the
expressions found in  eq. (A2) of Ref. \cite{PRA.2008.77.033832} and
eqs. (9) and (22) of Ref. \cite{MPLA.2004.19.2147}, respectively.

In order to investigate the role played by the nonminimal substitution
in \eqref{eq:oscillator} as well as its physical implications, we
should evaluate the nonrelativistic limit of eq. \eqref{eq:diracoquad}.
In this case, writing $E=M+\mathcal{E}$, with $M \gg \mathcal{E}$, we get
\begin{equation}\label{eq:diraconr}
  \left[
   \frac{p^2}{2M}+\frac{1}{2}M\omega^2r^2-\omega(\sigma^{3}+s L_{z})
  \right]\psi=
  \mathcal{E}\psi.
\end{equation}
The first two terms on the left side of eq. \eqref{eq:diraconr} are
those that appear in the Hamiltonian of the nonrelativistic circular
harmonic oscillator \cite{Book.1999.Flugge}, explaining why this system
is called Dirac oscillator.
The third term is a constant which shifts all energy levels.
The last term is the spin-orbit coupling, which (restoring the factor
$\hbar$) is of order $\omega/\hbar$. 
Summarizing, the nonrelativistic limit of the Dirac oscillator in $(2+1)$
dimensions is the circular harmonic oscillator with a strong spin-orbit
coupling term with all levels shifted by the factor $\omega$.
Indeed, the shifted energy levels are
\begin{equation}\label{eq:nrenergy2ddof}
\mathcal{E}+\omega=(1+2n+|m|-sm)\omega.
\end{equation}
As for the relativistic case, the infinity degeneracy is also present.

In summary, we have shown that the energy spectrum of the Dirac
oscillator in $(2+1)$ dimensions depends on the value of spin projection 
parameter $s$.  
On the other hand, it has also been shown that energy spectrum is degenerated
for all possible values of $sm$, a behavior not discussed before in the
literature.
Additionally, an isolated bound state solution, excluded from the
Sturm-Liouville problem, was discussed.

\acknowledgments
We would like to thanks to L. R. B. Castro for fruitful discussions. 
This work was supported by the 
Funda\c{c}\~{a}o Arauc\'{a}ria (Grants No. 205/2013 (PPP) and
No. 484/2014 (PQ)),
and the Conselho Nacional de Desenvolvimento
Cient\'{i}fico e Tecnol\'{o}gico (Grants No. 482015/2013-6
(Universal) and No. 306068/2013-3 (PQ)) and FAPEMA (Grant
No. 00845/13). 
Finally, we acknowledge some suggestions made by the anonymous referees
in order to improve the present work.

\bibliographystyle{eplbib}

\end{document}